\documentclass[prl,twocolumn,notitlepage,superscriptaddress,longbibliography]{revtex4-1}
\usepackage{amsmath}
\usepackage{amssymb}

\usepackage[unicode=true,colorlinks=true,citecolor=blue,urlcolor=blue]{hyperref}

\usepackage{bm}
\usepackage{color}
\usepackage{epsfig}
\usepackage{multirow}

\usepackage[normalem]{ulem}

\renewcommand {\Re}{\mathop\mathrm{Re}\nolimits}
\newcommand {\Tr}{\mathop\mathrm{Tr}\nolimits}

\renewcommand {\phi}{{\varphi}}
\newcommand {\rmi}{{\rm i}}

\newcommand {\e}{{\rm e}}

\newcommand{\rrangle}{\rangle\!\rangle}
\newcommand{\llangle}{\langle\!\langle}
\begin{document}
\title{%
Driven anti-Bragg subradiant states in waveguide quantum electrodynamics
}

\author{Alexander N. Poddubny}
\email{poddubny@coherent.ioffe.ru}

\affiliation{Ioffe Institute, St. Petersburg 194021, Russia}

\begin{abstract}
We  study theoretically driven quantum dynamics in periodic arrays of two-level qubits coupled to the waveguide.   We demonstrate, that strongly subradiant eigenstates of the master equation for the density matrix emerge under strong coherent driving for arrays with the anti-Bragg periods $d=\lambda/4,3\lambda/4$\ldots This happens even though are no such  states at low driving powers. Thus,  our findings show that in the quantum nonlinear regime the  domain of existence of subradiant states expands beyond conventional  subwavelength structures. These driven states are  directly manifested in long-living quantum correlations between the qubit excitations. 
\end{abstract}
\date{\today}

\maketitle
\begin{figure}[b]
\centering\includegraphics[width=0.45\textwidth]{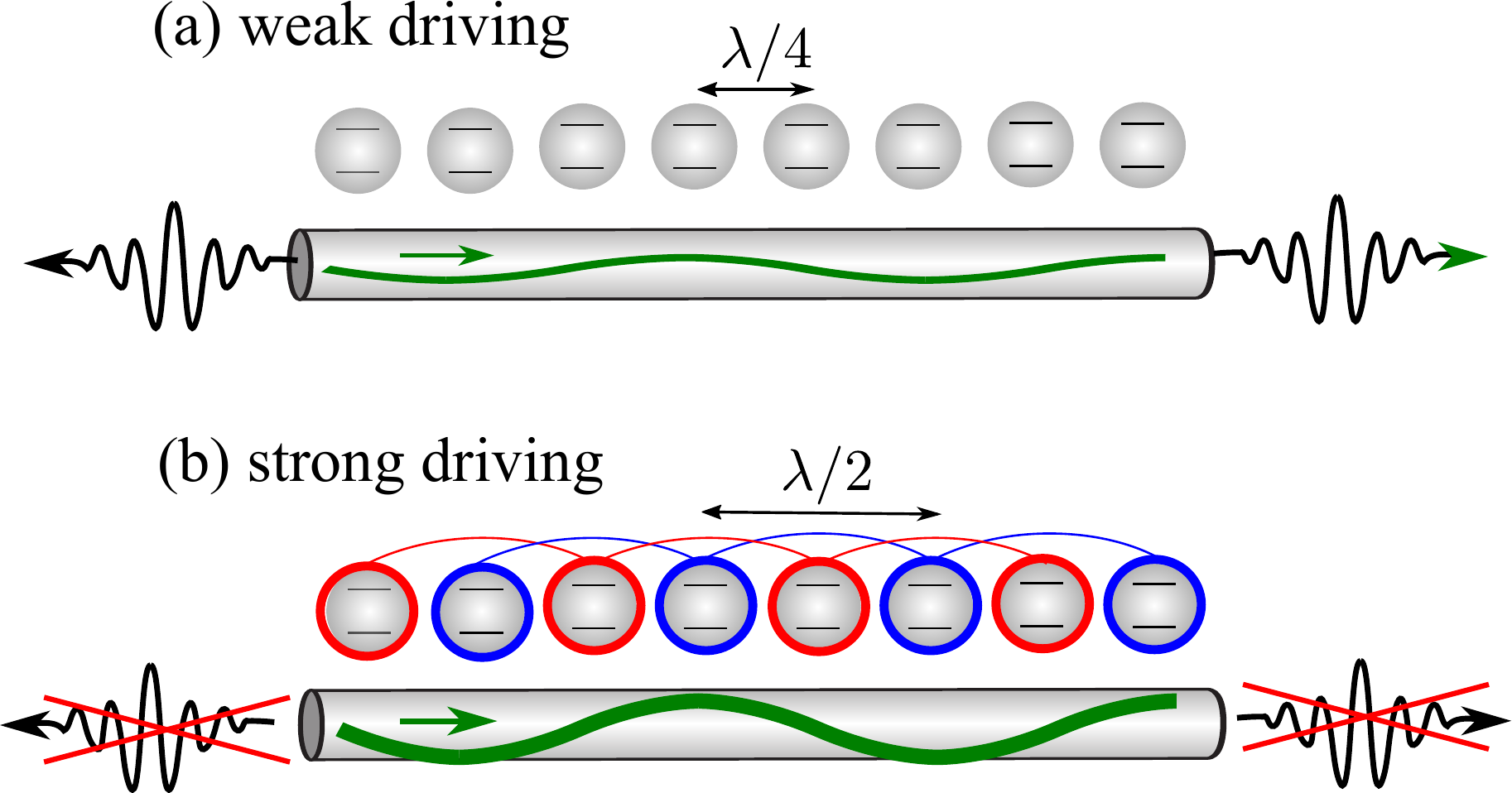}
\caption{Schematic illustration of an array of qubits coupled to a waveguide with an anti-Bragg period of $\lambda/4$. Panels (a) and (b) correspond to weak and strong coherent driving incident from the left side, respectively, illustrated by a green wave. In case (a) there are no subradiant  states in the system. In case (b) the array is split by driving into two $\lambda/2$-spaced subarrays and  subradiant states with suppressed spontaneous decay emerge in each subarray.}\label{fig:1}
\end{figure}

 
{\it Introduction.} 
Many-body quantum systems in the presence of driving and dissipation are now a subject of active studies.  It is now understood  that their dynamics is not limited to  conventional relaxation to the ground state and thermalization. Instead,  the system  can exhibit many-body localization~\cite{DAlessio2016,fayard2021manybody}, or even more exotic time-crystalline  phases that break the discrete time translational symmetry under the presence of external driving and oscillate in time instead of decaying to the time-independent stationary phase~\cite{Wilczek2012,Sacha2017,buca2019}. 
Here, we predict the formation of driving-induced nondecaying phases in the waveguide quantum electrodynamics (WQED) platform, shown in Fig.~\ref{fig:1}, where a periodic array of qubits is coupled to photons in a waveguide~\cite{Roy2017,KimbleRMP2018,sheremet2021waveguide}.

Suppression of spontaneous emission  by destructive interference is known at least since the work of Dicke~\cite{Dicke1954}. In a nutshell,  two quantum emitters can be excited with opposite phases to the state $|\psi\rangle=(\sigma_1^\dag-\sigma_2^\dag)|0\rangle/\sqrt{2}$ (here $\sigma_{1,2}^\dag$ are the  raising operators). Subradiant states have  been also extensively studied in the last several years for arrays of qubits coupled to the waveguide, both theoretically ~\cite{Molmer2019,Ke2019,Poddubny2019quasiflat,Zhang2020d} and experimentally~\cite{brehm2020waveguide,zanner2021coherent}. 
Nevertheless,  to the best of our knowledge, previously considered subradiant states are very distinct from those studied in this work because of their two following features.
First,   typical subradiant states exist only for relatively low filling factors of the array~\cite{Molmer2019}. Specifically, it has been proven in our recent work \cite{Poshakinskiy2021dimer} that many-body subradiant eigenstates of the effective non-Hermitian Hamiltonian of the array disappear for the fill factor above $f=1/2$ which indicated that they can be seen only under a relatively weak driving.  Indeed, increase of the driving strength typically results in the growth of linewidth and saturation of optical transitions of a two-level system~\cite{Astafiev2010}.
Second, a typical subradiant state  is a feature of an array with the subwavelength period ($d\ll \lambda$) or, more generally, a Bragg  period that is close to an any integer multiple of $\lambda/2$~\cite{ivchenko1994}.  On the other hand, for two qubits with the $\lambda/4$ spacing the waveguide-induced coupling has a purely exchange character:  both eigenmodes have the same lifetime as a single qubit and and neither super- no sub-radiant states exist. This has been experimentally demonstrated in Ref.~\cite{vanLoo2013}. It is however precisely this anti-Bragg regime with $d=\lambda/4, 3\lambda/4,\ldots$ that we focus on here. We predict, that while  conventional subradiant states do not exist for $d= \lambda/4$, they emerge  under strong coherent driving through the waveguide mode. Importantly, these states manifest  themselves only in the  full master equation for the density matrix of the driven system. This explains why they have not previously been revealed in the spectrum of effective many-body Hamiltonians analyzed in Refs.~\cite{Molmer2019,Ke2019,Poddubny2019quasiflat,Zhang2020d}.

%%%%%%%%%%%%%%%%%%%%%%%%%%%%%%%%%%%%%%%%%%%%%%%%%
%%%%%%%%%%%%%%%%%%%%%%%%%%%%%%%%%%%%%%%%%%%%%%%%%

The qualitative origin of such anti-Bragg driven subradiant states is illustrated in Fig.~\ref{fig:1}b.  Our main finding is that the strong coherent electromagnetic driving at the qubit resonance frequency leads to an effective splitting of the array into two  parts, the first part including only odd-numbered qubits and the second part with only even-numbered qubits. The driven $\lambda/4$-spaced array behaves akin to two $\lambda/2$-spaced subarrays. For $\lambda/2$-spaced arrays usual super- and sub-radiant states can form, e.g.$(\sigma_1^\dag+\sigma_3^\dag)|0\rangle/\sqrt{2}$ and they are revealed in the time dynamics. Thus, the driving breaks the spatial symmetry of the array and separates a non-subradiant system into two subradiant ones. Such symmetry breaking is a  mesoscopic feature of finite-size arrays that occurs even though  external driving is homogeneous in space, the array is periodic and has no disorder.

%\cite{Kozin2019}

{\it Model.}    The time dynamics of the density matrix  of the system $\rho$
%\begin{equation}\label{eq:Liouville}
 is described by the master equation  
$\partial_t \rho=\mathcal L\rho $
where $\mathcal L$ is the
 Lindblad superoperator, defined as \cite{Blais2013},
\begin{multline}
\mathcal L\rho=2\gamma_{\rm 1D}\sum\limits_{m,n=1}^{N}  \cos [\varphi(m-n)]
\sigma_{m}\rho \sigma_{n}^{\dag}-\rmi (H^\dag\rho-\rho H)\:.\label{eq:Lgen}
\end{multline}
Here, $\gamma_{\rm 1D}$ is the spontaneous  decay rate of a single qubit into the waveguide mode, $\omega_0$ is the qubit resonance frequency,  $c$ is the light speed in the waveguide and $\varphi\equiv 2\pi d/\lambda$  is the light phase gained between two neigboring qubits. The Hamiltonian of the structure reads
\begin{equation}\label{eq:H0}
H=H_0+V,\quad H_0=-\rmi \gamma_{\rm 1D}\sum\limits_{m,n=1}^N \sigma_m^\dag \sigma^{\vphantom{\dag}}_{n}\e^{\rmi \varphi|m-n|}\:,
\end{equation}
where the $H_0$ term describes the waveguide-induced coupling between the qubits. The non-Hermitian part of the Hamiltonian $H_0$ accounts for the spontaneous decay into the waveguide.
This Hamiltonian also assumes usual Markovian and rotating wave approximations. 
 The interaction term 
\begin{equation}\label{eq:V}
V=\Omega_{\rm R} \sum\limits_{n=1}^N(\sigma_n^\dag\e^{-\rmi \varphi n}+{\rm H.c.})
\end{equation}
is responsible for the resonant coherent driving at the qubit resonance frequency. For simplicity we count all the frequencies from the qubit resonant frequency $\omega_0$. 

\begin{figure}[t]
\centering\includegraphics[width=0.48\textwidth]{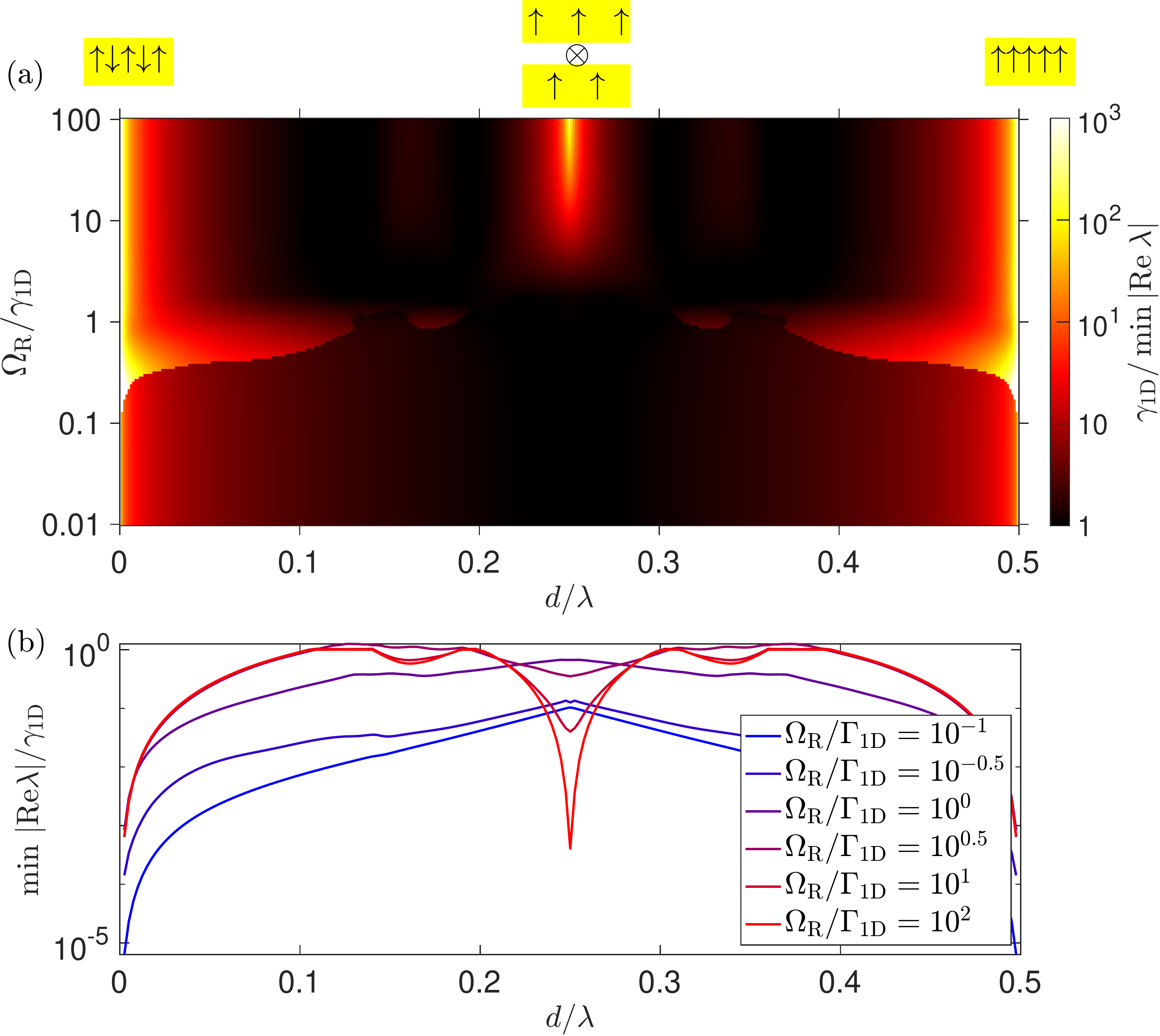}
\caption{ Dependence of the lifetime of the second longest-living state on the array period $d$ and the Rabi frequency $\Omega_{\rm R}$.  Panel(a) shows the color map, panel (b) shows the cross sections of the color map for several Rabi frequency values indicated on graph. The symbols on top of panel (a) schematically indicate the spin phases.
Calculation has been performed for $N=5$.}\label{fig:2}
\end{figure}

{\it Many-body time dynamics.}  
In order to provide insight into the time dynamics we study the eigenstates of the 
master equation defined as
%\begin{equation}\label{eq:lambda}
$\mathcal L\rho=\lambda\rho\:.$
We show in Fig.~\ref{fig:2} the dependence  of the spontaneous decay rate of the second longest living state $-\Re \lambda$ on the Rabi frequency $\Omega_{\rm R}$ and the array period $d$. We focus on the second-longest-living state since  there always is  one  trivial stationary eigenvalue of the Lindblad operator, exactly equal to zero,  that exists for any system parameters. Figure~\ref{fig:2}(a) shows the decay rate  as a color map, while Fig.~\ref{fig:2}(b) presents the cross section of the map vs. $d/\lambda$ for specific values of $\Omega_{\rm R}$.
The calculation demonstrates that at low powers, $\Omega_{\rm R}\ll \gamma_{\rm 1D}$, the  strongly subradiant states exist for $d$ close to $0$ and $d$ close to $\lambda(\omega_0)$. These states become less subradiant at larger powers, compare the first three curves in   Fig.~\ref{fig:2}(b) that are calculated for $\Omega_{\rm R}/\gamma_{\rm 1D}=10^{[-1,0.0.5,0]}$.

These results are  in full agreement with those in Refs.~\cite{Albrecht2019,Molmer2019,Ke2019,Zhang2020d,Poshakinskiy2021dimer}.
However, we stress that the methodology of the current study is conceptually different. These previous works, including ours \cite{Ke2019,Poshakinskiy2021dimer} ,   have been focused  on the eigenstates of the non-Hermitian Hamiltonian $H_0$ in Eq.~\eqref{eq:H0}, that are characterized by a certain integer number of polaritons $\hat n=\sum_m \sigma_m^\dag \sigma_m^{\vphantom{\dag}}$. On the other hand, here we consider the eigenvalues of the total master equation in the presence of driving. Hence, the current study captures additional physics that is manifested at larger powers, $\Omega_{\rm R}\gg \gamma_{\rm 1D}$, that is beyond the effective Hamiltonian approach, and, to the best of our knowledge, has not been analyzed before.
%%%%%%%%%%%%%%%%%%%%%%%%%%%%%%%%%%%%%%%%%%%%%%%%%

Specifically, it can be seen in Fig.~\ref{fig:2} that while strongly   subradiant states are 
not present for   an anti-Bragg period $d\approx \lambda(\omega_0)/4$ under weak  driving, they emerge for $\Omega_{\rm R}\gg\gamma_{\rm 1D}$. This is also manifested as a sharp dip in the middle of the last two red curves in  Fig.~\ref{fig:2}(b). The larger is power the longer is the lifetime of these states.  An observation of these  anti-Bragg subradiant states at larger powers  constitutes our main finding. 

\begin{figure}[t]
\centering\includegraphics[width=0.4\textwidth]{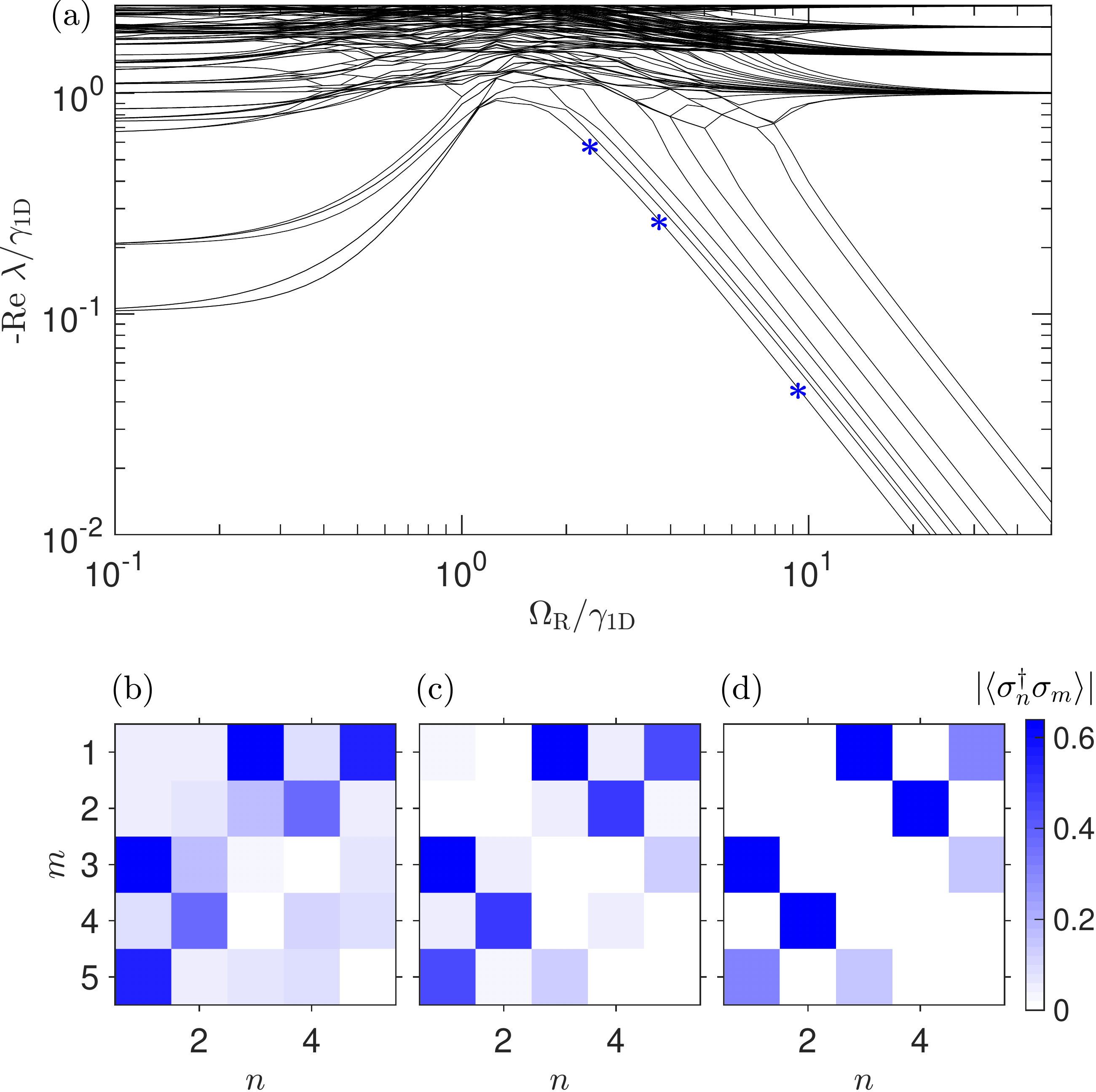}
\caption{(a) Dependence of the normalized decay rates on $\Omega_{\rm R}$  for $d=\lambda/4$ and $N=5$. (b,c,d) Spin-spin correlation function $|\langle \sigma_n^\dag \sigma_m\rangle|$ calculated for the second longest-living state for $\Omega_{\rm R}/\gamma_{\rm 1D}=10^{0.4},10^{0.6},10^1$. These values are also indicated by blue stars  in panel (a). }\label{fig:power}
\end{figure}

In order to examine the  anti-Bragg subradiant states in more detail we show in  Fig.~\ref{fig:power}
the dependence of the decay rates on $\Omega_{\rm R}$ for $d=\lambda/4$.
At low powers,  $\Omega_{\rm R}\ll \gamma_{\rm 1D}$, there are no strongly  subradiant states in the system. It can be shown that the decay rate of the longest living states for $d=\lambda/4$ ($\varphi=\pi/2$) is on the order of $\Re \lambda=-\pi^2\gamma_{\rm 1D}/N^3$, in contrast to the much smaller decay rate 
$\Re \lambda\sim \varphi^2\gamma_{\rm 1D}/N^3$ in the subwavelength structures where $\varphi\ll \pi$~\cite{Molmer2019,brehm2020waveguide}.   At larger powers, $\Omega_{\rm R}\lesssim \gamma_{\rm 1D}$, these decay rates become even larger, which can be viewed as a result of a saturation of the qubit transitions induced by the driving. However, the spectrum drastically changes  for $\Omega_R\gtrsim 10\gamma_{\rm 1D}$.  Nine eigenvalues split from the rest of the spectrum, and acquire small real parts $\sim \gamma_{\rm 1D}^3/\Omega_{\rm R}^2$ that decrease at larger powers. These is the manifestation of considered  anti-Bragg subradiant states.

The spatial structure of these states is further examined in Fig.~\ref{fig:power}(b--d) where we present the 
correlation function $\Tr [\rho  \sigma_n^\dag \sigma_m] $ calculated for the second longest-living state at three different powers. With the increase of the driving  the correlations acquire a characteristic checkerboard pattern, namely, they are present only between the qubits of the same parity.
This observation suggests that the arrays with odd- and even-numbers qubit, that have $\lambda/2$ spacing [see also Fig.~\ref{fig:1}b], should be analyzed  separately at larger powers.
%%%%%%%%%%%%%%%%%%%%%%%%%%%%%%%%%%%%%%%%%%%%%%%%%
\begin{table}[b]
\begin{tabular}{|c|c|c|c|c|c|}
\hline
&  $N_{\rm even}$&1 & 2 & 3&4\\\hline
$N_{\rm odd}$ & $N_{\rm dark}$&1&2&5&14\\
\hline
$1$&1&$\times$&2&$\times$&$\times$\\
$2$&2&2&4&10&$\times$\\
$3$&5&$\times$&10&25&70\\
$4$&14&$\times$&$\times$&70&196\\\hline
\end{tabular}
\caption{Number of dark eigenstates of the master equation in subarrays of even- and odd-numbered qubits with $d=\lambda/2$ and in the total anti-Bragg array with $d=\lambda/4$ and $N=N_{\rm even}+N_{\rm odd}$. More details are given in text.}\label{table:1}
\end{table}

{\it Odd- and even- numbered subarrays.} We start this analysis by noting, that according to the calculation in Fig.~\ref{fig:2}, the subradiant eigenstates of the master equation partially survive at high powers not only for the anti-Bragg period, but also for the subwavelength period $d\ll\lambda(\omega_0)$ and the Bragg period $d\approx \lambda(\omega_0)/2$. This is best shown by the red curves of  Fig.~\ref{fig:2}b that presents the dependence  of the decay rates on period for  fixed large Rabi frequencies: there are distinct minima for $d=0$ and $\lambda/2$. These states  appear for $N\ge 2$ qubits and correspond  to the product of usual single-particle subradiant states of the effective Hamiltonian Eq.~\eqref{eq:H0}. For example, for $N=2$ one has
\begin{equation}\label{eq:rhodark}
 \rho=|\psi\rangle\langle\psi|,\quad |\psi\rangle=\frac{1}{\sqrt{2}}(\sigma_1^\dag\mp\sigma_2^\dag)|0\rangle
\end{equation}
with antiferromagnetic (ferromagnetic) coupling for $d=0$ and $d=\lambda/2$, respectively. It can be directly seen that the density matrix in Eq.~\eqref{eq:rhodark} commutes with the driving operator $V\equiv\Omega_{R}(\sigma_1\pm \sigma_2)+{\rm H.c.}$. Thus, the state Eq.~\eqref{eq:rhodark} is not affected by the driving. Since it is dark at low driving powers it remains dark at all values of $\Omega_{\rm R}$. There also is one other trivial eigenstate for $N=2$ corresponding to $\rho=1$. Using the same logic, it is possible to find 5 dark eigenstates of the master equation for $N=3$ qubits with the $d=0$ or $d=\lambda/2$ spacing. The correspond to a trivial eigenstate, $\rho=1$, and four different products of two dark eigenstates
$(\sigma_1^\dag\mp\sigma_2^\dag)|0\rangle$ and $(\sigma_1^\dag\mp\sigma_3^\dag)|0\rangle$. For $N=4$ there exist $14$ subradiant eigenstates of the master equation and so on.

We  have numerically studied the number of subradiant eigenstates of the master equation in the anti-Bragg structures with up to $8$ qubits. The results are summarized in Table~\ref{table:1}. We also present there the number of the dark eigenstates for each even- and odd-numbered subarray with $\lambda/2$ spacing. For example, the table cell with $N_{\rm even}=2$ and $N_{\rm odd}=3$ corresponds to the array with $N\equiv N_{\rm even}+N_{\rm odd}=5$. Such array has ten subradiant states, while its subarrays have two and five subradiant states, respectively. Since $10=2\times 5$, there is a strong hint that the eigenstates of the total array correspond just to product of the eigenstates of each subarrays. This is indeed confirmed by the results in Table~\ref{table:1} and also agrees with  Fig.~\ref{fig:power}. We illustrate the antiferromagnetic and ferromagnetic structure of  subradiant states at $d=0,\lambda/2$ as well as the product structure of the states at $d=\lambda/4$ by the up and down spin strings  at the top of Fig.~\ref{fig:2}.

%%%%%%%%%%%%%%%%%%%%%%%%%%%%%%%%%%%%%%%%%%%%%%%%%
%%%%%%%%%%%%%%%%%%%%%%%%%%%%%%%%%%%%%%%%%%%%%%%%%
%%%%%%%%%%%%%%%%%%%%%%%%%%%%%%%%%%%%%%%%%%%%%%%%%

%%%%%%%%%%%%%%%%%%%%%%%%%%%%%%%%%%%%%%%%%%%%%%%%%
\begin{figure}[b]
\centering\includegraphics[width=0.4\textwidth]{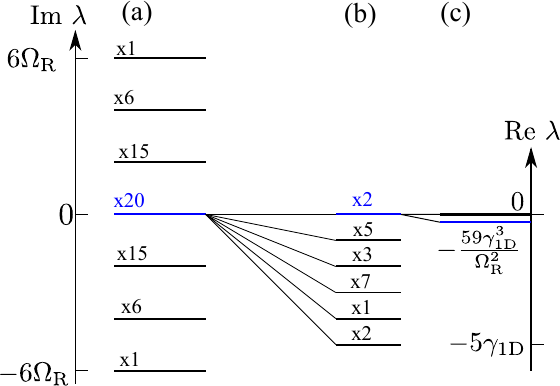}
\caption{Scheme of the splitting of the eigenvalues of the Linblad equation in zeroth(a), first (b) and third (b) order in the parameter $\gamma_{\rm 1D}/\Omega_{\rm R}$, that leads to the formation of a many-body subradiant state.}\label{fig:levels}
\end{figure}
%%%%%%%%%%%%%%%%%%%%%%%%%%%%%%%%%%%%%%%
%%%%%%%%%%%%%%%%%%%%%%%%%%%%%%%%%%%%%%%%%%%%%%%%%

%%%%%%%%%%%%%%%%%%%%%%%%%%%%%%%%%
\begin{figure}[t]
\centering\includegraphics[width=0.46\textwidth]{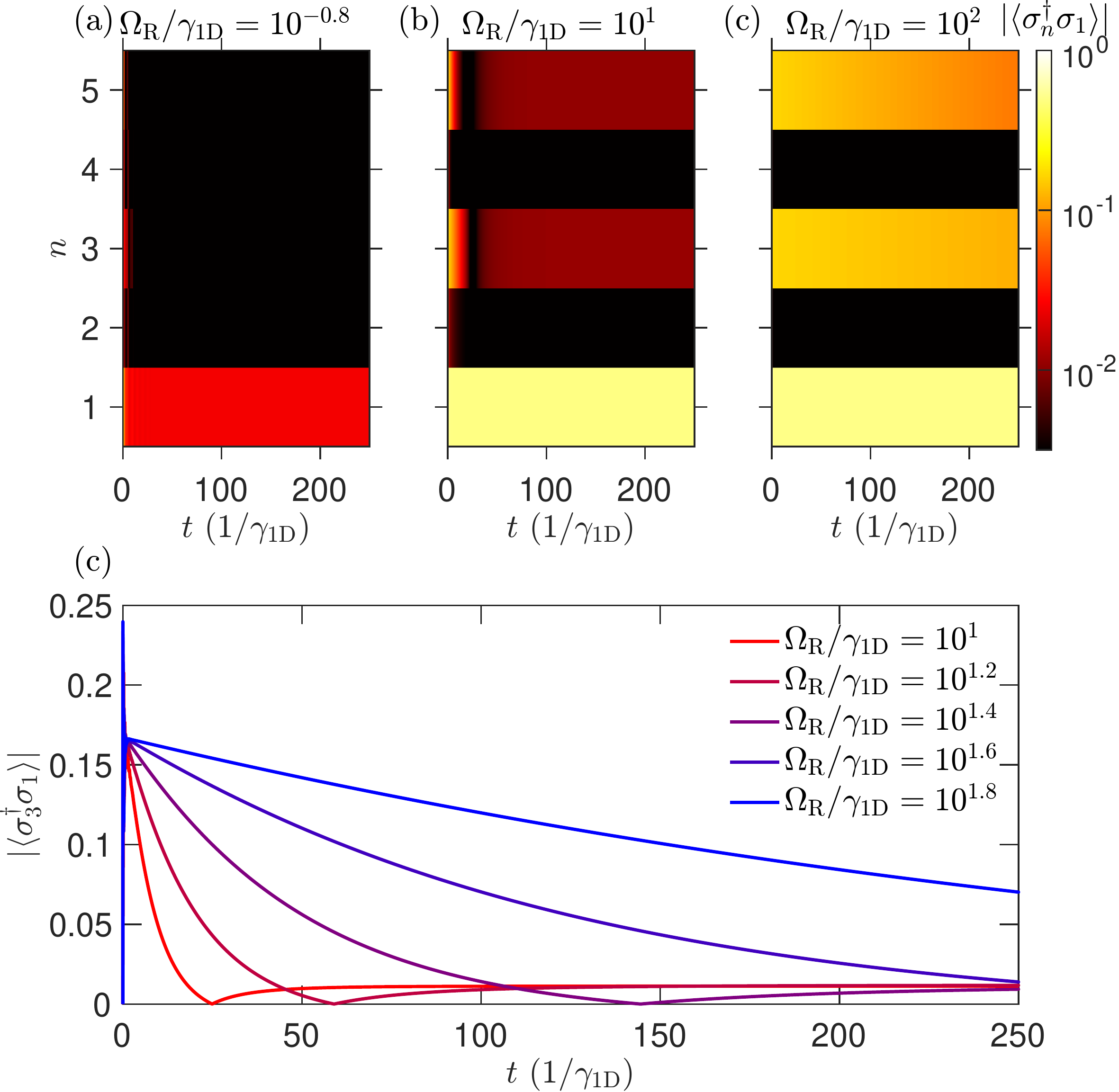}
\caption{Time dynamics of the correlation function $\langle \sigma_n^\dag \sigma_1 \rangle$ depending on the driving power.
(a--c) Dependence of the correlation  functions for three different driving strength, indicated on graphs.
(d) Time dependence  of the correlation function $\langle \sigma_3^\dag \sigma_1 \rangle$.
Calculation has been performed for $d=\lambda/4, N=5$. At the moment $t=0$ the array is in the fully  excited state. }\label{fig:4}
\end{figure}
%%%%%%%%%%%%%%%%%%%%%%%%%%%%%%%%%

%%%%%%%%%%%%%%%%%%%%%%%%%%%%%%%%%
{\it Perturbation theory}.
%%%%%%%%%%%%%%%%%%%%%%%%%%%%%%%%%
A more rigorous description of the anti-Bragg subradiant states can be obtained by using the perturbation theory in the master equation. To this end we split the Lindblad superoperator \eqref{eq:Lgen}
 as $\mathcal L=\mathcal L_0+\mathcal L_V$, with  
$ \mathcal L_V\rho\equiv\rmi [\rho,V]$. The eigenstates of the $\mathcal L_V$ operator, responsible for the driving,  can be easily found analytically. Next, the effect of the waveguide-mediated coupling and dissipation, governed by the operator $\mathcal L_0$, is included as a perturbation series in a small parameter $\gamma_{\rm 1D}/\Omega_{\rm R}$. 

For $\lambda/4$ spacing the $V$ operator Eq.~\eqref{eq:V} describes just an interaction  of $N$ spins-1/2 with the magnetic field $\Omega_R$, that periodically rotates in space along $z$ axis with the period of 4. After the basis is appropriately rotated, 
$\sigma_m^\dag\e^{-\rmi \varphi m}\to \widetilde\sigma_m^\dag$, the eigenstates $|J_z\rangle$ of the $V$ operator can  be characterized by the projection of the total angular momentum $J_z$. For example, for $N=3$ qubits there exist two non-degenerate states with the eigenvalues $\lambda_{J_z}=\pm 3\Omega_{\rm R}$ ($J_z=\pm 3/2$) and two triplets with the eigenvalues $\lambda_{J_z}=\pm \Omega_{\rm R}$ ($J_z=\pm 1/2$). The 64 eigenstates of the operator $\mathcal L_V$ are formed as 
$\rho_{J_{z},J_{z}'}\equiv |J_{z}\rangle\langle J_{z}' |$\:, with the eigenvalues $\lambda_V= -2\rmi \Omega_R (J_z-J_{z}')$.
The distribution of the eigenvalues $\lambda_V$ is shown in Fig.~\ref{fig:levels}(a). It is straightforward to show that there are $1+3^2+3^2+1=20$ degenerate eigenvalues $\lambda_V=0$, corresponding to all combinations of pairs with $J_z=J_z'$. We will label these 20 zero eigenvalues as $|\nu\rrangle$ and all the rest eigenstates of the master equation as as $\rho^{\mu}\equiv |\mu\rrangle$, where $\mu$ runs from 1 to 44, with the eigenvalues $\lambda_{V,\mu}$. Next, we take into account the operator $\mathcal L_0$ as a perturbation in the third order, following the general approach from Ref.~\cite{BirPikus}:
\begin{align}\label{eq:L0}
\widetilde{\mathcal L_0}&=\mathcal P_0 \Bigr[\mathcal L_0\\&+  \mathcal L_0 G \mathcal L_0 \nonumber\\&+
  \mathcal L_0 G \mathcal L_0  G \mathcal L_0-\frac{
 \mathcal L_0 G^2  \mathcal L_0 \mathcal P_0\mathcal L_0+
  \mathcal L_0 \mathcal P_0\mathcal L_0 G^2 \mathcal L_0 }{2}\Bigr]\mathcal P_0\:,\nonumber
\end{align}
where $\mathcal P_0\equiv \sum_\nu |\nu\rrangle\llangle \nu| $ is the projector on the subspace of the states with $\lambda_V=0$ and the term
%\begin{equation}
$G=-\sum_{\mu=1}^{44 }\frac1{\lambda_\mu}|\mu\rrangle\llangle\mu|$
%\end{equation}
describes the admixture of the states with $\lambda_{V,\mu}\ne 0.$
Each line in Eq.~\eqref{eq:L0} corresponds to a certain perturbation order in $\gamma_{\rm 1D}/\Omega_{\rm R}$, from the first order to the third  order. The result of the spectrum splitting in the first order is shown in Fig.~\ref{fig:levels}(b). In this order there exist two degenerate eigenstates with $\lambda_V=0$ corresponding to the density matrices
\begin{gather}\label{eq:rho12}
\rho_1=1,\text{ and }\rho_2\propto 4|\psi_2\rangle \langle \psi_2| \otimes |\psi_{1,3}\rangle\langle\psi_{1,3}|-1,\\
|\psi_{1,3}\rangle=\frac{1}{\sqrt{2}}[\sigma_1^\dag |0\rangle+\sigma_3^\dag |0\rangle]\:,
|\psi_{2}\rangle=\sigma_2^\dag |0\rangle\:.\nonumber
\end{gather}
The state $\rho_1$ is just a trivial eigenstate with the unity matrix. The first term in the state $\rho_2$ also has a transparent interpretation as a described above pure dark state of odd-numbered qubits 1 and 3, and the unity is subtracted to account for orthogonalisation. 
 The spontaneous decay appears only in the second order in $(\gamma_{\rm 1D}/\Omega_{\rm R})^2$, as shown in Fig.~\ref{fig:levels}(c). The terms in the last line of Eq.~\eqref{eq:L0} lead to the splitting of the doublet. As a result, the state $\rho_2$ acquires a non-zero decay rate
 $\lambda=-\xi\gamma_{\rm 1D}^2/\Omega_R^3$. The coefficient $\xi$ is found to be equal to $59/9$ for $N=3$. The same procedure can be also performed for a larger number of qubits, yielding the numbers of subradiant states agreeing with those listed in Table~\ref{table:1}.

%%%%%%%%%%%%%%%%%%%%%%%%%%%%%%%%%%%%%%%%%%%%%%%%%
{\it Potential experimental observation.}
The  long-living eigenstates of the master equation can be directly observed by measuring the time-dependent  correlation functions between the qubit excitations. The results are shown in Fig.~\ref{fig:4} for different  driving strengths. We start with the array being fully excited at the moment $t=0$ and then study the evolution of the correlations $\langle\sigma_{n}^\dag\sigma_1\rangle$ depending on the driving strength. For low strength, Fig.~\ref{fig:4}(a) and red curves in Fig.~\ref{fig:4}(d), the correlations for $n>1$ quickly decay with time. An increase of the driving strength leads to appearance of the significant correlations 
$\langle\sigma_{3}^\dag\sigma_1\rangle$ and $\langle\sigma_{5}^\dag\sigma_1\rangle$ at large times. The blue curves in Fig.~\ref{fig:4}(d) clearly indicate the slowdown of the decay. This is fully consistent with the structure shown in Fig.~\ref{fig:power}. A potential experiment would require a superconducting processor with just $N=3$ anti-Bragg-spaced qubits,  which is well within the current technological limits of circuit QED~\cite{Blais2020}.

%%%%%%%%%%%%%%%%%%%%%%%%%%%%%%%%%%%%%%%%%%%%%%%%%

{\it Outlook.}   
Our findings uncover yet another mechanism of formation of subradiant states in the Dicke-like models by showing, that such states are not limited to subwavelength or Bragg-spaced structures. It might be  instructive to examine, whether the concept  of driving-induced-subradiance could be extended to other geometries, such as two-dimensional quantum metasurfaces~\cite{Rui2020}. Considered states could be probably also viewed as a quantum nonlinear generalization of the bound states in continuum~\cite{Hsu2016}. While the obtained states are degenerate, it might be also interesting to look for non-degenerate non-decaying states in the WQED setup, for example, in the presence of disorder, that could be  potentially related to nonstationary time-crystalline phases. The waveguide-mediated interactions are inherently long-ranged. This is beneficial for time-crystalline phases~\cite{Kozin2019}, but  has not been explored so far to the best of our knowledge.

\begin{acknowledgements}
I am  grateful to A.V.~Poshakinskiy for multiple useful discussions.
\end{acknowledgements}

%%%%%%%%%%%%%%
%\nocite{apsrev41Control}
%\bibliographystyle{apsrev4}
%\bibliography{titleon,timecryst}

\end{document}